\documentclass[%
 aip,
rsi,%
 amsmath,amssymb,
 reprint,%
]{revtex4-1}

\usepackage{graphicx}
\usepackage{dcolumn}
\usepackage{bm}
\usepackage{amsmath}
\usepackage{siunitx}
\usepackage{braket}
\usepackage{upgreek}

\usepackage{lipsum}

    \makeatletter
\def\@fnsymbol#1{\ensuremath{\ifcase#1\or \dagger\or \ddagger\or
   \mathsection\or \mathparagraph\or \|\or **\or \dagger\dagger
   \or \ddagger\ddagger \else\@ctrerr\fi}}
    \makeatother
    
\begin{document}


\title{Broadband pumped polarization entangled photon-pair source in a linear beam displacement interferometer}

\author{Alexander Lohrmann}
\email{cqtal@nus.edu.sg}
\affiliation{%
Centre for Quantum Technologies, National University of Singapore, 3 Science Drive 2, S117543, Singapore\\
}%

\author{Chithrabhanu Perumangatt}
\affiliation{%
Centre for Quantum Technologies, National University of Singapore, 3 Science Drive 2, S117543, Singapore\\
}%

\author{Aitor Villar}
\affiliation{%
Centre for Quantum Technologies, National University of Singapore, 3 Science Drive 2, S117543, Singapore\\
}%

\author{Alexander Ling}
\affiliation{%
Centre for Quantum Technologies, National University of Singapore, 3 Science Drive 2, S117543, Singapore\\
}%
\affiliation{%
Physics Department, National University of Singapore, 2 Science Drive 3, S117542, Singapore
}%



\begin{abstract}
We experimentally demonstrate a source of polarization entangled photon-pairs based on a single periodically-poled potassium titanyl phosphate (PPKTP) crystal pumped with a broadband, free running laser diode. The crystal is placed within a linear beam-displacement interferometer, and emits photon-pairs based on type-0 spontaneous parametric downconversion (SPDC). We observe pair rates of 0.56~Mpairs/s/mW in a single spatial mode with a polarization visibility of 97.7\% over a spectral range of 100~nm. This experiment demonstrates a pathway towards observing Gigacount rates of polarization entangled photon pairs by using high-power free-running laser diodes with fast multiplexed detectors. 
\end{abstract}


\maketitle
Photonic entanglement is a critical resource in many quantum technologies such as quantum key distribution, teleportation, metrology and secure time-transfer \cite{E91, teleport2, fink19, lee19}.
The building block in most of these technologies is the entangled photon-pair.
One of the most convenient methods of producing entangled photon-pairs is SPDC in birefringent crystals such as beta barium borate (BBO)\cite{kiess93,kwiat1995new}, potassium titanyl phosphate (KTP)\cite{pelton04} and lithium niobate\cite{tanzili01}.
As new use-cases of photonic entanglement are developed, and more experiments and applications outside the laboratory are planned\cite{yin2017satellite,oi17,wengerowsky18}, existing source designs might be found wanting in areas such as brightness, size or robustness, spurring the exploration of new geometries and approaches\cite{fiorentino2007spontaneous,trojek2008collinear,steinlechner2012high,steinlechner2013phase,steinlechner2014efficient, steinlechner2017distribution,chen2018polarization,villar2018experimental,lohrmann2018high, jabir2017robust}.

In terms of the source performance, two of the most vital parameters are the detected pair rate (observed brightness) and the entanglement quality of the photon-pairs. 
Significant effort has been made in the pursuit of maximizing those two parameters, from domain engineering to access higher nonlinear coefficients \cite{qpmshg} and correspondingly higher SPDC efficiency, to the design of different optical configurations to access different degrees-of-freedom \cite{barreiro2005generation} or to pursue size reduction and increased stability \cite{trojek2008collinear,villar2018experimental,lohrmann2018high,horn19}.

The most straightforward way to increase the observed brightness of SPDC-based entangled photon sources is to increase the pump power; the flux of downconverted photons being proportional to the input power. 
In recent years, the output power of free-running laser diodes has increased substantially while the cost per device has fallen.
While the output power of narrow linewidth pump sources have also increased, they typically lag behind free-running systems due to sensitivity to mode-hops and are correspondingly more challenging to operate.
Free-running blue laser diodes used to create photon-pairs in the near-infrared have linewidths that are typically hundreds of pico-meters ($\approx$ THz) wide.
This is not an issue from phase-matching considerations given that the SPDC process can accept a fairly broad pump spectrum.
However, the birefringent nature of the crystals, and other components introduce wavelength-related phase effects which will render the generated photon-pairs to be distinguishable in the production process.
Correspondingly, this can result in a deterioration of entanglement quality unless additional steps are introduced.
A common mitigation technique is to employ narrow-band interference filters in order to achieve acceptable entanglement quality \cite{steinlechner2014efficient, chen2018polarization}, which translates into a decreased brightness. Other configurations have been examined to enable the use of low cost broadband laser diodes resulting in relatively low brightness\cite{jeong2016bright}. 



In this work we present a source of high-quality polarization-entangled photon-pairs, based on a single periodically poled crystal that is pumped using a free-running laser diode. 
The resulting bandwidth of the SPDC emission spans over 100~nm, and high quality entanglement is achieved over this bandwidth by the combination of several elements such as pump compensation and the use of momentum correlation to split the photon-pair.
This source can achieve detected rates comparable to the brightest reports in the literature\cite{steinlechner2014efficient, steinlechner2013phase}. 
The class of free-running laser diodes that is used in the present work can reach an output power of up to 1~W and our design could be combined with multiplexed, high-throughput single photon detectors\cite{holzmann2019superconducting} to enable observation of unprecedented rates of entangled photon-pairs.

\begin{figure}
    \centering
    \includegraphics[scale = 0.8]{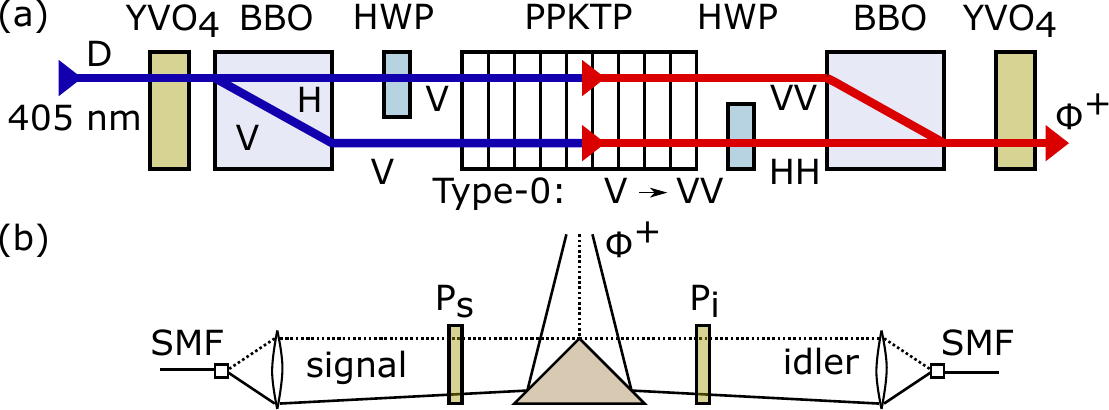}
    \caption{(a) Schematic of the main source components. The diagonally polarized pump is focused into the center of the PPKTP crystal with a waist size of $\omega_p=100$~\SI{}{\micro\metre} and a separation between the two paths of 1~mm. (b) Schematic of the splitting and detection setup. The photon-pair is split by its momentum components on the wedge mirror and the polarizers, $P_s$ and $P_i$ can be inserted to analyze the polarization state. Finally, the beam is coupled into single mode fibers and detected using Geiger mode avalanche photo diodes.
    }
    \label{fig:setup}
\end{figure}

We use a beam displacement inteferometer in which downconversion is generated along two separate paths within the same periodically poled crystal\cite{fiorentino2008compact}. 
The two paths are combined to generate the entangled state. 
The source concept is schematically shown in Fig.~\ref{fig:setup}.

The starting point of the source is a diagonally polarized pump beam ($\lambda_p = 405$~nm) that is split into its two polarization components, $\ket{H}$ and $\ket{V}$, using spatial walk-off in a birefringent crystal (BBO, length: 13 mm, cut-angle: 45$^\circ$). The resulting displacement between the two polarization components is 1~mm. The $\ket{H}$ component is rotated to $\ket{V}$ by a half-wave plate. Type-0 SPDC ($\ket{V}\rightarrow \ket{VV}$) is generated in both paths in the periodically poled crystal (PPKTP, length: 10~mm, poling period: \SI{3.425}{\micro\metre}). After downconversion one of the SPDC beams is rotated from $\ket{VV}$ to $\ket{HH}$ using another HWP. Finally, another BBO crystal (length: 13.76~mm, cut-angle: 45$^\circ$) combines the two beams, generating the entangled state:

\begin{equation*}
\ket{\psi} = \frac{1}{\sqrt{2}}\left( \ket{HH} + e^{i\Delta\varphi}\ket{VV} \right),
\end{equation*}

\noindent where $\Delta\varphi$ denotes the phase difference between $\ket{HH}$ and $\ket{VV}$. The emission wavelength can be tuned from degenerate (810~nm) to non-degenerate emission by temperature tuning the PPKTP crystal.

A wedge mirror is employed to separate the photon-pair according to its momentum correlation. This has the advantage that it does not introduce a wavelength dependent phase, as is often found when splitting using dichroic mirrors. With this method ``signal" and ``idler" comprise the entire spectral region.

To flatten the phase that is introduced by the wavelength dependent birefringence in the walkoff crystals we employ pre- and post-compensation with additional birefringent crystals\cite{trojek2008collinear, rangarajan2009optimizing,villar2018experimental}. Yttrium vanadate (YVO$_4$) was selected owing to its high birefringence and dispersion. Optimal compensation is achieved when an a-cut YVO$_4$ crystal with a length of \SI{0.78}{mm} is placed in the pump beam, and another YVO$_4$ crystal with a length of \SI{0.97}{mm} is placed in the combined SPDC beam (see Supplementary Information).

\begin{figure}
    \centering
    \includegraphics{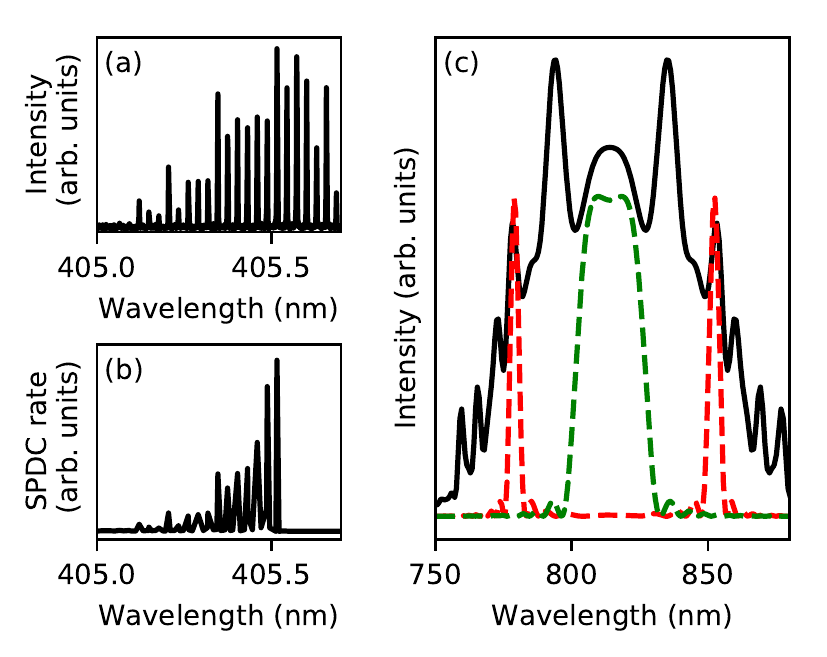}
    \caption{(a) Typical spectrum of the free-running laser diode used in the present work. The available output power of the laser diode is 150~mW. (b) Collinear SPDC rate into a single spatial mode as a function of pump wavelengths weighted with the spectrum of the pump spectral modes. When the degenerate point is exceeded, the SPDC emission becomes non-collinear and we assume that the collected pair rate is negligible. (c) Expected SPDC emission spectrum based on the pump distribution in (a) (black line) and two spectra assuming a narrowband pump (green: degenerate, red: non-degenerate). Inspection of the curves will reveal that the spectral features for both cases are slightly broader at the longer wavelengths. }
    \label{fig:specCalc}
\end{figure}

An important consideration when using a broadband pump is the sensitivity of collinear type-0 phasematching to the pump wavelength. 
One strategy to achieve maximum brightness when using a broadband pump is to align the peak of the pump emission modes (see Fig.~\ref{fig:specCalc}(a)) to the collinear degenerate phase matching condition. 
The other pump frequency modes may trigger non-degenerate SPDC, but at lower efficiency. 
The SPDC output as a function of the pump wavelength is shown in Fig.~\ref{fig:specCalc}(b), and the corresponding spectrum is shown in Fig.~\ref{fig:specCalc}(c). 
The spectrum stretches over a spectral range from 760 to 870~nm. For comparison, we have included the SPDC spectra generated for an ideal narrowband pump laser.

This source configuration has a small foot print and the interferometer spanned by the walk-off crystals does not require alignment. 
The source employs two possible decay paths in a single direction for the pump photon while using only a single crystal. This results in reduced operational requirements compared to designs employing either two crystals \cite{steinlechner2012high,horn19} or back-reflection \cite{steinlechner2013phase}.



\begin{figure}
    \centering
    \includegraphics{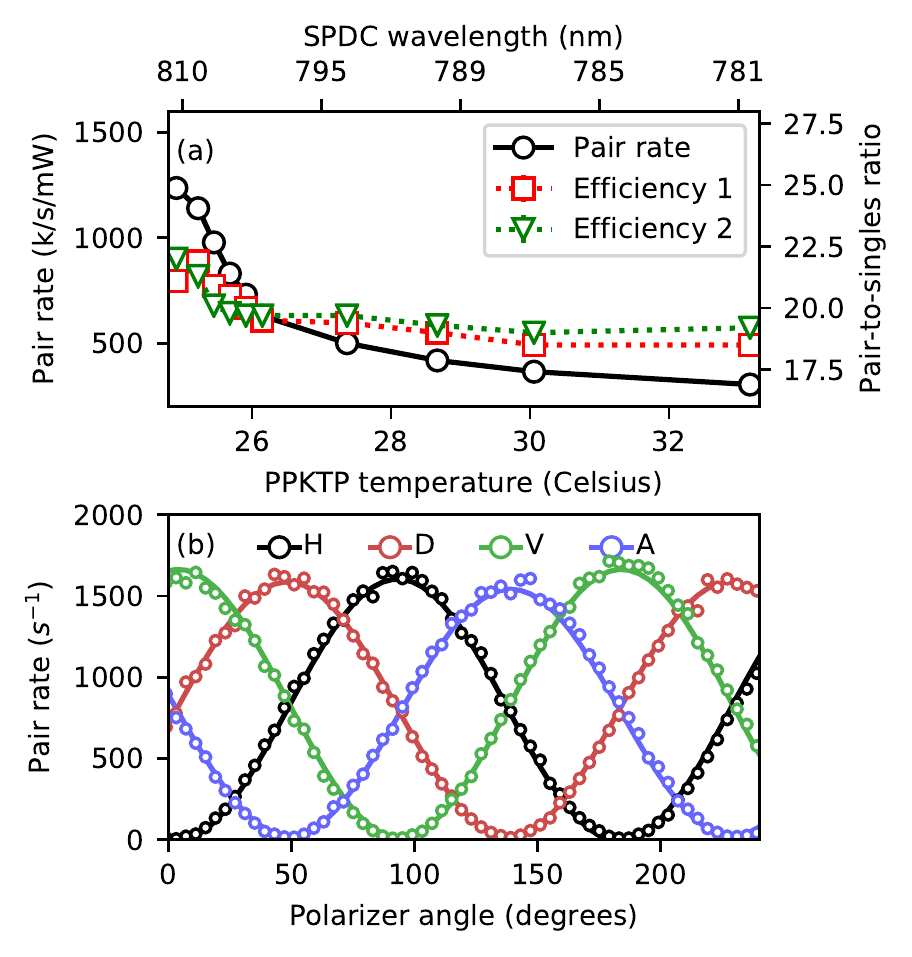}
    \caption{(a) Pair rate (black circles) and pair-to-singles ratio (red squares and green triangles) as a function of PPKTP crystal temperature when using a narrowband pump. The detected pair rate close to the degenerate edge reaches 1.25 Mpairs/s/mW. (b) Raw observed pair rate as a function of polarizer angle in the signal arm for four fixed polarizer settings in the idler arm. The input power was \SI{5}{\micro\watt} and the PPKTP crystal temperature was approximately $26^\circ$~C. Note that the polarizers that were used in the measurement have transmission efficiency of 85\% which slightly reduces the measured pair rate. This visibility can be achieved across the full temperature/wavelength range.}
    \label{fig:ondaxCurves}
\end{figure}

The source described above was first prepared using a narrowband laser diode ($\Delta\nu < 160$~MHz) as the pump. 
The intention in this step was to demonstrate that the assembled components could indeed produce high-quality entanglement, and to serve as a benchmark when the free-running laser was used later.
To avoid saturation of the passively quenched single photon detectors, we use a suitably low input power $P < 100$ \SI{}{\micro\watt}. 
Due to manufacturing restrictions, the lengths of the pre- and post- compensation crystals were 0.92~mm and 1.04~mm, respectively (see Supplementary Material). This results in a less than optimal phase map, but the effect on the entangled state is not significant, as will be shown.

The detected pair rate and pair-to-singles ratio (an estimate of collection efficiency) as a function of the crystal temperature, are shown in Fig.~\ref{fig:ondaxCurves}(a). 
The observed pair rate reaches a normalized value of 1.2~Mpairs/s/mW for degenerate emission and drops to 0.3~Mpairs/s/mW for the wavelengths of 780/842~nm. 
The corresponding pair-to-singles ratio drops monotonically from 22\% to 19\% over this temperature range. 
The spectral width of the SPDC emission is 14~nm for the degenerate case and drops below 3~nm for the non-degenerate emission around 780/842~nm\cite{steinlechner2014efficient}.
This data suggests that the brightness and pair-to-singles ratio would be acceptably high when using the broadband pump.

Fig.~\ref{fig:ondaxCurves}(b) shows the typical non-local polarization correlation for four polarizer settings (H/D/V/A) without correction for the residual accidental coincidences. 
The visibility in the H/V basis reaches $99.2\pm0.2$\% while the visibility in D/A reaches $98.4\pm0.2$\%.
This visibility can be achieved for SPDC wavelengths across the spectral range from 780~nm to 842~nm. 

\begin{figure}
    \centering
    \includegraphics{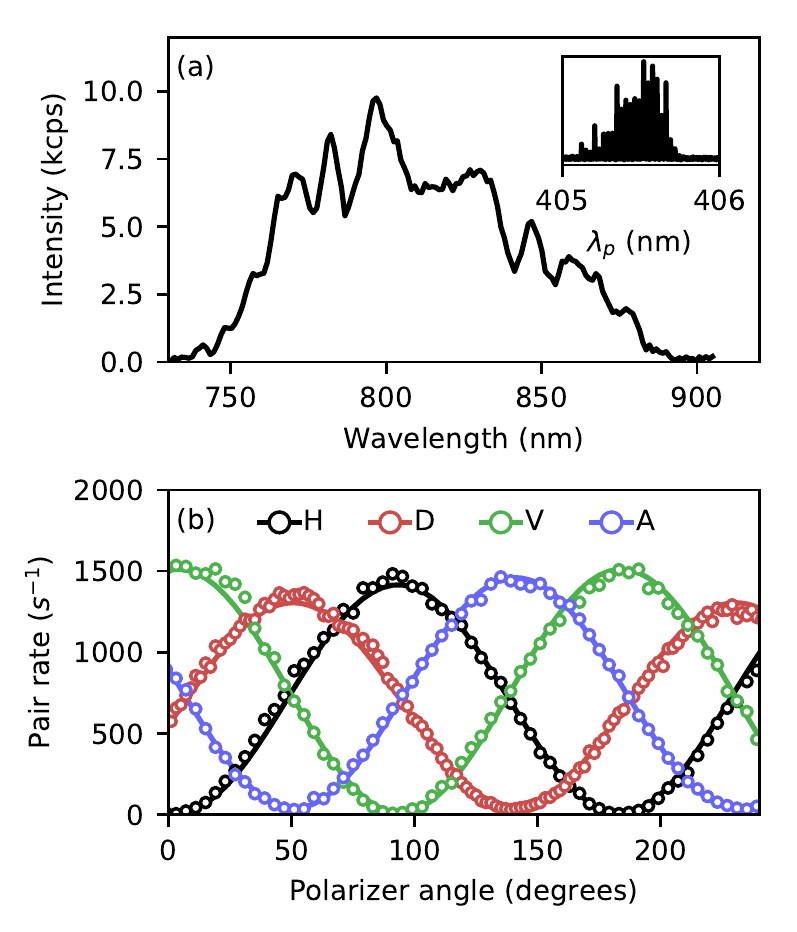}
    \caption{(a) SPDC spectrum measured for the broadband pumped source. The observed spectrum lacks symmetry compared to Fig.~\ref{fig:specCalc}(c) as the spectrometer and single photon detectors have a wavelength dependent efficiency. A typical emission spectrum of the pump is shown in the inset. (b) Uncorrected pair rate as a function of polarizer angle in the signal arm for four fixed polarizer settings in the idler arm. The input power was approximately \SI{10}{\micro\watt}. Compared to Fig.~\ref{fig:ondaxCurves} the normalized brightness per unit of pump power is lower due to the challenge of collecting the broad emission into a single spatial mode using refractive lenses.}
    \label{fig:broadbandCurves}
\end{figure}

After introducing a free-running laser diode as the broadband pump, the SPDC emission is distributed over a spectral range of approximately 100~nm, as shown in Fig.~\ref{fig:broadbandCurves}(a). The measured pair rate reached 0.560~Mpairs/s/mW with a pair-to-singles ratio of 21\% for both signal and idler. 
This pair rate is in good agreement with the brightness when using a narrowband pump, if the broader pump and emission spectrum are taken into account. 

The corresponding non-local polarization correlation is shown in Fig.~\ref{fig:broadbandCurves}(b). 
Without correcting for accidental coincidences, the visibility of the broadband SPDC source is $ 99.0\pm0.2$\% in the H/V basis, while the visibility in D/A reaches $96.4\pm0.4$\%. The lower visibility is attributed to the non-ideal lengths of the compensation crystal (see Supplementary Information). 
Despite this, an average intrinsic source visibility of 97.7\% was observed which is suitable for many applications, e.g. in quantum key distribution it would account for an intrinsic quantum bit error rate of approximately 1\%.

The use of the broadband pump has several advantages. 
First, free-running laser diodes have an extremely high power-to-cost ratio when compared with narrowband lasers, making them cost-effective for pumping the SPDC process. Furthermore, laser diodes with an emission at 405~nm and a bandwidth of 0.5~nm can easily provide an output power of 1~W. Using such laser diodes with this design can permit rates of 10~billion pairs per second to be generated in a single spatial mode. 
The challenge in observing such a high rate is in the lack of fast single photon detectors.
However, it is possible to multiplex the emission in space or wavelength into multiple channels. The multiplexing effectively distributes the detected pair rates across different detectors, while maintaining the strong wavelength correlations. This allows to overcome the fundamental limitation of multi-pair emission, since the accidental pair rate is proportional to the product of the single event rates (see Supplementary Information). 

Multiplexing the pairs by wavelength makes use of the wavelength correlations between signal and idler photons that are inherent to the SPDC emission process. For broadband pumped SPDC, the wavelength correlation of each pair depend on the wavelength of the parent pump photon. Therefore, a mismatch between the wavelength correlations of photon pairs originating from different pump photons could result in a case where one pair is multiplexed in the correct channels, whereas a pair pumped by a different wavelength is not. This could in principle reduce the multiplexing efficiency. However, since type-0 phasematching efficiency itself is critically dependent on the pump wavelength (see Fig.~\ref{fig:specCalc}(c)), photon pairs generated by different pump wavelengths have only minimal spectral overlap. Therefore, multiplexing is valid despite the use of a broadband pump.


In the context of terrestrial quantum key distribution multiplexing can increase the total secret key rate, and opens the possibility for multi-user networks\cite{lim2008broadband,kang2016monolithic,chapuran2009optical,wengerowsky18_1}. Another interesting application involves quantum communication under high-loss, as to be expected in fundamental tests of quantum physics from deep space\cite{cao2018bell} or entanglement distribution from geostationary orbits.





Overall, the source combines a small footprint with ruggedness against misalignment.
If necessary, the design can be further simplified by combining displacement and pre-compensation crystals.
The source design is flexible and can be adjusted for highly non-degenerate emission and different pump and SPDC wavelengths. 
In conclusion, this source provides a realistic path for upgrading the pump power, increasing the feasibility of detection (as opposed to production) of entangled photon-pairs at Gigacounts per second in a single spatial mode, which is conceivably important for future quantum networking applications.

\section*{Supplementary Material}
See supplementary material for a more detailed motivation of the mutliplexing scheme and information on the phase compensation crystals.

\section*{Funding}
National Research Foundation (NRF); Competitive Research Programme (CRP); Award No. NRF-CRP12-2013-02; Ministry of Education (MOE); Research Centres of Excellence (RCE).

\bibliography{bib}

\end{document}